\documentclass[aps,twocolumn,prl,showpacs,color,psfig,epsf]{revtex4}

\usepackage{amsmath}
\usepackage{amsfonts}
\usepackage{graphicx}
\usepackage{hyperref}

\hypersetup{
    colorlinks=true,
    linkcolor=blue,
    citecolor=blue,
    filecolor=blue,
    urlcolor=blue,
}

\makeatletter
\newcommand*{\balancecolsandclearpage}{%
  \close@column@grid
  \clearpage
  \twocolumngrid
}
\makeatother

\newcounter{suppequation}
\newcounter{suppfig}
\makeatletter
\@addtoreset{equation}{suppequation}
\@addtoreset{figure}{suppfig}
\makeatother

\begin{document}

\title{Intracellular facilitated diffusion: searchers, crowders and blockers}

\author{C. A. Brackley, M. E. Cates, D. Marenduzzo}
\affiliation{SUPA, School of Physics and Astronomy, University of Edinburgh, Mayfield Road, Edinburgh, EH9 3JZ, UK}

\begin{abstract} 
In bacteria, regulatory proteins search for a specific DNA binding target via ``facilitated diffusion": a series of rounds of 3D diffusion in the cytoplasm, and 1D linear diffusion along the DNA contour. Using large scale Brownian dynamics simulations we find that each of these steps is affected differently by crowding proteins, which can either be bound to the DNA acting as a road block to the 1D diffusion, or freely diffusing in the cytoplasm. Macromolecular crowding can strongly affect mechanistic features such as the balance between 3D and 1D diffusion, but leads to surprising robustness of the total search time.
\end{abstract}

\maketitle

Many cellular processes involve the binding of a protein to a specific target base pair (bp) sequence of DNA~\cite{alberts}. As targets are typically 10-20 bp in size, it is remarkable that proteins can quickly and accurately locate them on a bacterial chromosome which is over $10^6$ bp long. Interest in the mechanism of the target search first arose in the 1970s, when experiments by Riggs et al.~\cite{Riggs1970} measured the \emph{in vitro} rate at which the \emph{lac} repressor binds with its promoter; early interpretations of these results found that rate to be much greater than would be expected for a search via simple diffusion in three dimensions~\cite{fn:riggs}. This prompted a series of papers by Berg and von Hippel~\cite{Berg1981,vonHippel1989}, in which they developed an analytical model of a protein-DNA target search known as {\it facilitated diffusion}. Their premise was that portions of the search involve events where, via a non-specific (sequence independent) interaction the searching protein ``slides'' along the DNA backbone --- effectively performing diffusion in one dimension. Following these seminal works much theoretical and computational effort~\cite{Sokolov2005,Sokolov2005-2,Loverdo2009,Zabet2012,Stormo1998,Slutsky2004,Gerland2002,Benichou2009,Sheinman2012,Marcovitz2011,Brackley2012,Florescu2009,Florescu2010} has been spent addressing different facets of this important problem.

However, the process is still not fully understood, despite recent advances in single molecule imaging techniques that have given new insight into facilitated diffusion mechanisms~\cite{Elf2007,Wang2006,Hammar2012}. Most importantly, the bacterial cytoplasm within which the search is performed is very different from the test tube of {\it in vitro} experiments: it is a very crowded environment, with about a million proteins per cell~\cite{alberts,Pelletier2012}, many of which bind to the DNA and perform functions such as transcription, gene regulation, replication and repair, and chromosome structuring~\cite{alberts,Nicodemi}. So far, only limited theoretical~\cite{Li2009,Mirny2009} or simulation work explicitly addresses this issue. Furthermore, colloidal physics arguments suggest that crowding should give rise to depletion~\cite{Asakura1958} and other entropic interactions between the constituents of the system: these are important for the thermodynamics of many intracellular processes~\cite{Marenduzzo2006,verkman}, but their impact on facilitated diffusion has not yet been explored.

Our goal in this Letter is to investigate the effects of macromolecular crowding, both in the cytosol and along the DNA, on the target search process. Our most interesting results are those for diffusing crowder proteins. These strongly bias the search towards 1D diffusion along the genome, as found \emph{in vivo}~\cite{Elf2007}, and through an intriguing combination of effects lead, at physiological crowder densities, to a search time which is robust to changes in protein-DNA affinity, as might be beneficial within a living cell. We also show that blockers (proteins tightly bound to, and diffusing along, the DNA) do not greatly hinder facilitated diffusion, even at surprisingly large densities. To gain our results, we use coarse grained Brownian dynamics (BD) simulations. These allow us to explicitly include the (often disregarded) conformational dynamics of DNA; provide us with a detailed treatment of the 3D component of the search; and avoid the need to make the \emph{a priori} assumptions on the relative timescales of the kinetics in the search process which are necessary in most analytical approaches~\cite{Berg1981,Halford2004,Slutsky2004}.

\begin{figure}[t]
\includegraphics{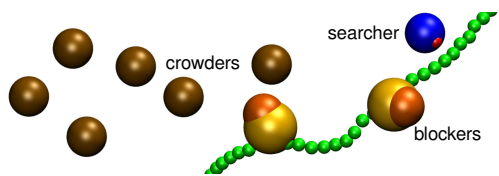}
\caption{A bead-and-spring DNA with \emph{crowder}, \emph{blocker} and \emph{searcher} proteins. \label{picture} }
\end{figure}

For the simulations we coarse grain DNA as a bead-and-spring polymer. The level of coarse graining we use (the size of a polymeric bead equals the hydration thickness of DNA, $2.5~\mbox{nm}$) allows many repeat simulations of large systems to be performed, at the cost of losing atomistic details in the resolution of the genome and protein structures. As in Ref.~\cite{Brackley2012} we model searcher proteins as a sphere with a small patch; the latter is sticky and acts as a DNA binding site, interacting with an energy $\epsilon$. For simplicity, we neglect sequence heterogeneity -- this can potentially lead to more complicated 1D dynamics which have been studied in Ref.~\cite{Brackley2012} in the absence of macromolecular crowding~\cite{notedisorder}. We introduce two further protein types which we shall refer to as \emph{crowders} and \emph{blockers} [Fig.~\ref{picture}]. Crowders are proteins which diffuse freely in the space around the DNA, and are modelled as simple spheres. Blockers are modelled as larger proteins which bind strongly to the DNA (but can still diffuse along it), blocking the path of the searchers from all sides. The simulations are performed using the LAMMPS code~\cite{lammps}. We use a system size which reproduces a typical bacterial DNA volume fraction of 1\%; periodic boundary conditions are employed, so that the genome is under compaction (its radius of gyration is larger than the system size), similar to the situation \emph{in vivo}. Full details of the simulation scheme are given in Ref.~\cite{SI}.

We analyse our results with reference to the analytic model of Halford and Marko~\cite{Halford2004}: whilst more detailed models exist~\cite{Benichou2009,Sheinman2012,cherstvy,kolomeisky}, this framework allows a simple explanation of how different parameters affect the search. According to Ref.~\cite{Halford2004}, if during each 1D sliding episode the protein searches on average a length $l_s$, then it will  require on average $N_s=L/l_s$ rounds of 3D and 1D diffusion to find the target on a DNA molecule of length $L$. This relationship between $N_s$ and $l_s$ depends on the assumption that there is no correlation between the regions of the DNA which are searched during successive slides (some recent theories have sought to address this~\cite{Benichou2009,cherstvy,kolomeisky}). Scaling arguments then give the relations $\tau_{\rm 3D}\sim V/LD_3$ and $\tau_{\rm 1D}\sim l_s^2/D_1$ for the mean duration of each 1D or 3D search event, where $D_3$ and $D_1$ are 3D and 1D diffusion constants, and $V$ is the system volume. This leads to an equation $\tau=N_s(\tau_{\rm 1D}+\tau_{\rm 3D})$, and finally
\begin{equation}
\tau=A\frac{L l_s}{D_1} + B\frac{V}{D_3l_s}, \label{eq:theory}
\end{equation}
for the mean total search time. Here the dimensionless prefactors $A,B$ cannot be inferred from simple scaling. 

\begin{figure}[t]
\includegraphics{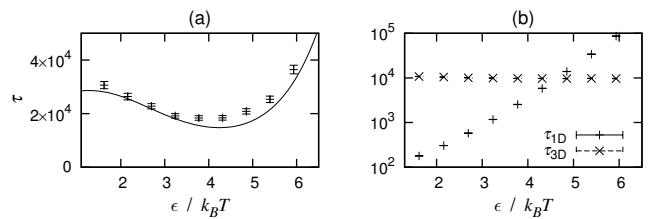}
\caption{Facilitated diffusion in the absence of macromolecular crowding. (a) Plot showing how the total search time depends on the searcher-DNA interaction strength $\epsilon$, for a system with a single searcher protein, and no blockers or crowders. The solid line shows a fit to Eq.~(\ref{eq:theory}). (b) Plot showing the mean duration of 1D and 3D search episodes from the same simulations. \label{fig1} }
\end{figure}

In Ref.~\cite{Brackley2012} we showed that Eq.~(\ref{eq:theory}) only gives a good fit to the BD results for an \emph{unstructured} DNA molecule. Fig.~\ref{fig1}(a) shows the mean total search time as a function of the protein-DNA interaction $\epsilon$ for such a system, and Fig.~\ref{fig1}(b) shows individually the 1D and 3D search times. In the theory of Ref.~\cite{Halford2004} $l_s$ is taken as a parameter, whereas in the BD model we can only directly vary $\epsilon$. Now $l_s$ is a function of $\epsilon$, as is $D_1$ (which was not the case in~\cite{Halford2004}, but might be expected for a protein sliding through a rugged potential due to a sequence of discrete monomers). This means that there is an optimal value of $\epsilon$ where the search time is minimised. From Eq.~(\ref{eq:theory}) at the minimum the fraction of time spent sliding $f=\tau_{1D}/(\tau_{1D}+\tau_{3D})=1/2$. (This holds only approximately in Fig.~\ref{fig1} due to the additional dependence of $D_1$ on $\epsilon$ in our simulations.) 

A natural starting point to address the effects of crowding would be to consider a system containing \emph{multiple searcher proteins} and a DNA molecule with a single target. If the search for the target were a simple Poisson process, one would expect the mean search time to scale inversely with the number of searchers, i.e. the search time for $M$ searchers should obey $\tau(M)=\tau(1)M^{-1}$. This relationship gives a very good approximation to our simulation results in most cases, and we provide full details in~\cite{SI}.

\begin{figure}
\includegraphics{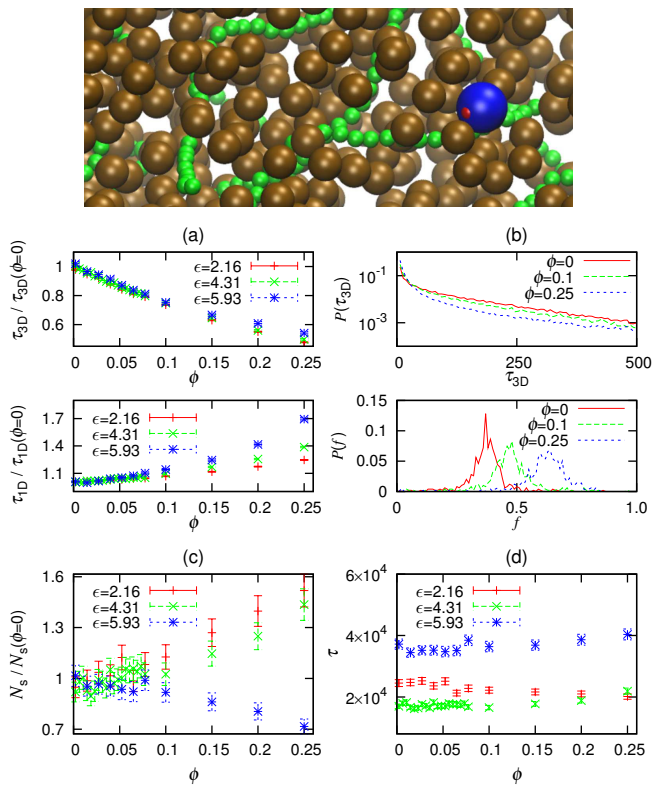}
\caption{Simulation results for systems containing crowder proteins, and a single searcher. Top: Snapshot of the DNA, crowders and the searcher during the simulation. (a) The mean 3D and 1D search times scaled by the time for $\phi=0$; (b) the probability distribution functions for the duration of 3D excursions (top), and for the fraction of time spent sliding, $f$ (bottom), for $\epsilon=4.31~k_BT$;  (c) the mean number of search rounds required to find the target, scaled by the value for $\phi=0$; and (d) the  mean total search time. \label{fig:crowders} }
\end{figure}

Much more interesting consequences arise when considering the search process in the presence of many \emph{crowder proteins}, whose volume fraction we denote by $\phi$. The presence of the crowders has a large, but opposite effect on the 3D and 1D components of the search [Fig.~\ref{fig:crowders}(a)]: an increase in crowder density from zero to $\phi=0.25$ leads to a roughly two-fold decrease in $\tau_{\rm 3D}$, but to an increase by a similar factor in $\tau_{\rm 1D}$. Both results can be attributed to a depletion-like interaction \cite{Asakura1958,Marenduzzo2006} between the searcher and the DNA: when the searcher is close to the DNA, there is a region between it and the polymer from which crowders are excluded due to steric interactions. The resulting osmotic pressure due to the crowders acts to effectively increase the searcher-DNA attraction when the protein is bound (a known effect of macromolecular crowding \cite{Zimmerman87}), and an increased likelihood of the searcher immediately returning to the DNA after is has escaped. This manifests as the progressive deviation away from a simple exponential for the probability distribution of 3D excursion times [Fig.~\ref{fig:crowders}(b), top], in favour of a distribution suggestive of biphasic or stretched exponential behavior. 

Interestingly, the effect of increasing crowder density on the number of search rounds $N_s$ is also different depending on the value of $\epsilon$ [Fig.~\ref{fig:crowders}(c)]. There are two competing effects here. Due to the increase of $\tau_{1D}$, we expect an increase of $l_s$ with $\phi$ which should lead to a decrease in $N_s$ (assuming $N_s= L/l_s$). However the diffusion of both the DNA and the searcher~\cite{fn:diffusion} is hindered by the crowders and this leads to an increased likelihood of repeatedly searching the same DNA region, causing a breakdown of the $N_s= L/l_s$ relation. This is demonstrated by the fact that the product $N_sl_s$ increases with $\phi$ for all $\epsilon$ (see~\cite{SI}). $N_s$ increases with $\phi$ for small $\epsilon$, but for the $\epsilon=5.93~k_BT$ case shown in the figure, the increase in $l_s$ dominates, and $N_s$ \emph{decreases} with $\phi$.

The various dependencies on $\phi$ of $N_s$, $\tau_{\rm 1D}$ and $\tau_{\rm 3D}$ at different $\epsilon$ combine to give remarkable stability of the total mean search time $\tau$ [Fig.~\ref{fig:crowders}(d)]. For small $\epsilon$, $\tau$ decreases with $\phi$ whereas for large $\epsilon$ it increases. However, despite the large changes in $\tau_{\rm 1D}$  and $\tau_{\rm 3D}$, the overall change in $\tau$ remains small (see~\cite{SI}) . Moreover, at the physiologically relevant $\phi\approx0.2$, $\tau$ shows little variation between the $\epsilon\sim 2~k_BT$, and $4~k_BT$ cases (a plot showing $\tau$ as a function of $\epsilon$ is given in~\cite{SI}).

The depletion attraction thus enhances the robustness of the search time. According to Eq.~(\ref{eq:theory}) the cell must precisely tune the protein-DNA interaction to optimise the search --- difficult, since $\epsilon$ will vary from protein to protein, depend on DNA sequence, and be dependent on conditions such as salt concentration. Thus our finding that $\tau$ is relatively insensitive to $\epsilon$ near its minimum suggests that the facilitated diffusion search mechanism is better adapted to the subcellular conditions in which it operates than to a hypothetical, uncrowded environment. Another important result is that there is not a clear link between the fraction of time spent sliding, $f$ (recently measured \emph{in vivo}~\cite{Elf2007}), and the search time. From Eq.~(\ref{eq:theory}), an optimal search leads to $f=0.5$, whereas Fig.~\ref{fig:crowders}(b) shows that the probability distribution of $f$ gets shifted to sliding-dominated searches as $\phi$ increases, independent of $\epsilon$, with $f$ increasing twofold as $\phi$ increases from $0$ to $0.25$.

\begin{figure}
\includegraphics{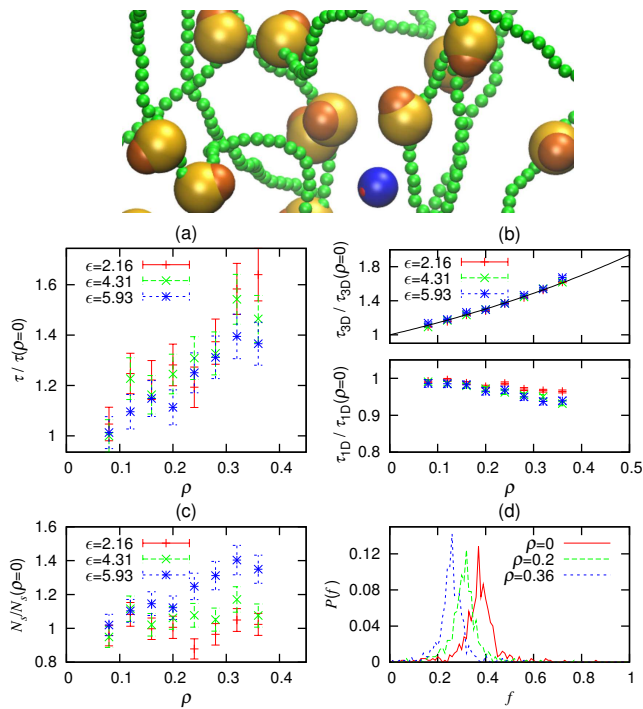}
\caption{Simulation results for systems containing blocker proteins, and a single searcher; $\rho$ is defined as the average fraction of DNA covered by blockers during the simulation. Top: Snapshot of the DNA, blockers and the searcher during the simulation. (a) The mean search time as a function of $\rho$, scaled by the search time for the $\rho=0$ case. (b) The mean 3D and 1D search times as a function of $\rho$. Black line shows an exponential fit to the $\epsilon=4.31~k_BT$ results. (c) The mean number of search rounds required to find the target, scaled by the value for $\rho=0$.  
(d) The probability distribution function for the fraction of time spent sliding, $f$, for $\epsilon=4.31~k_BT$.\label{fig:blockers} }
\end{figure}

We finally consider \emph{blocker proteins}, which bind to, and diffuse along, the DNA rather than diffuse within the cytosol. For transparency we consider their effects in isolation, addressing a system with no crowders and a single searcher. Similar ``road-block" proteins have previously been considered analytically in~\cite{Li2009,Mirny2009}. The theory by Li \emph{et al.}~\cite{Li2009} predicts an overall increase in the search time $\tau$ due to the reduced $l_s$, and the chance that the target is covered by a blocker. Our results are qualitatively in agreement with this, showing an increase in $\tau$ with average blocker coverage density $\rho$ which is almost independent of $\epsilon$ [Fig.~\ref{fig:blockers}(a)]. 

Quantitatively, however, this increase is not dramatic; this is perhaps another well-adapted feature of facilitated diffusion within a cell where a large number of DNA-binding proteins must be present at any time~\cite{alberts}. We find that the duration of each 3D excursion ($\tau_{\rm 3D}$) increases with $\rho$ since it takes longer for the searcher to encounter DNA which is free from blockers; Fig.~\ref{fig:blockers}(b) top  shows a good fit to an exponential function, which is expected since the probability of finding an uncovered region of a given length decreases exponentially with coverage density \cite{Li2009,McGhee1974}. In contrast to Ref.~\cite{Li2009}, we also observe a small decrease in $\tau_{\rm 1D}$ as $\rho$ increases [Fig.~\ref{fig:blockers}(b), bottom]. Presumably, collisions between the searcher and the blockers can lead to the less strongly bound searcher being ``knocked off'' of the DNA. We note that in our model, searchers can only bypass blockers by transiently detaching from the DNA. While this is a good assumption for large blocking proteins or protein clusters (such as polymerases, replication or transcription factories), a searcher may in practice hop over smaller proteins without leaving the genome, as discussed in Ref.~\cite{Marcovitz2013} -- this effect may further diminish the slowdown of the search process imparted by blockers.

Fig.~\ref{fig:blockers}(c) shows that the number of search rounds also increase with $\rho$. This stems from the reduction in $l_s$ (due to reduction in $\tau_{1D}$), and of the increased likelihood of the target being covered by a blocker. Due to the effects in Fig.~\ref{fig:blockers}(b), the presence of blockers also affects the fraction of time spent sliding, $f$: now the target-finding mechanism becomes more and more dominated by 3D excursion as the DNA gets increasingly obstructed [Fig.~\ref{fig:blockers}(d)]. 

To summarize, in this work we have studied the effect on facilitated diffusion of macromolecular crowding caused by the simultaneous presence of many proteins in the cellular environment. Crowding proteins diffusing in the cytosol lead to depletion interactions which effectively increase the searcher-DNA affinity, and reduce the 3D search time, at the same time increasing the chance of repeat-pass searches through the same DNA region. This leads to a decrease in search time for small $\epsilon$ but an increase for large $\epsilon$ meaning that at physiological volume fractions of crowders the curve shown in Fig.~\ref{fig1}(a) becomes flatter, with $\tau$ increasing only at large $\epsilon$. That $\tau$ is independent of $\epsilon$ near its minimum may be advantageous for efficient searching, since in reality the non-specific searcher-DNA interaction must vary both among searchers and along the DNA sequence. Robustness to changes in $\epsilon$ has previously been shown to result from intersegmental transfers~\cite{fninterseg,Sheinmann2009}, in that case at high DNA-protein affinity; this process relies on DNA strands passing close to each other in space and so depends on the compaction of the DNA. Due to their geometry, intersegmental transfers occur relatively infrequently with our searchers; concurrent investigation of both of these effects would be an interesting future study.

For DNA-binding \emph{blocker} proteins we have confirmed previous analytical work which predicted an exponential increase in the 3D search time due to impaired access to the DNA~\cite{Li2009}. Our simulations also suggest that the blocker-induced increase in the total search time is modest, which is again advantageous in the biological context where a large number of DNA-binding proteins must be present simultaneously for the proper functioning of both bacterial and eukaryotic chromosomes. Finally, we have shown that if multiple searcher proteins are present then the search time scales inversely with copy number, except at very high copy numbers (see~\cite{SI}).

Our work suggests that a proper account of the crowded cellular environment is crucial for a full understanding of protein-DNA target search. This is becoming an increasingly important issue as direct experimental probes of the process \emph{in vivo} become a reality~\cite{Elf2007,Hammar2012}. We have found that, at least in terms of the robustness of the total search time $\tau$, the facilitated diffusion mechanism is well-adapted to the crowded subcellular conditions in which it takes place. This work also lays the foundations for future studies of the combined effect of crowders and blockers; the role of architectural DNA-binding proteins; the effect of genome confinement~\cite{Foffano2012,Bauer2013}; and the possible influence of hydrodynamic interactions~\cite{Dunweg2009}. 

\begin{acknowledgments}
We acknowledge EPSRC grant EP/I034661/1 for funding. MEC is funded by the Royal Society.
\end{acknowledgments}

\balancecolsandclearpage
\newpage

\renewcommand{\thefigure}{S\arabic{figure}}
\renewcommand{\theequation}{S\arabic{equation}}

\stepcounter{suppequation}
\stepcounter{suppfig}


\renewcommand*{\citenumfont}[1]{S#1}
\renewcommand*{\bibnumfmt}[1]{[S#1]}

\onecolumngrid
\section*{\Large Intracellular facilitated diffusion: searchers, crowders and blockers.}

\section*{\Large Supplemental Material}
\begin{center}
C. A. Brackley, M. E. Cates, D. Marenduzzo \\
\textit{SUPA, School of Physics and Astronomy, University of 
Edinburgh, Mayfield Road, Edinburgh, EH9 3JZ, UK}
\end{center}

\vspace{1cm}

\twocolumngrid

\section*{Methods}

As noted in the text, for the simulations we coarse grain DNA as a bead-and-spring polymer, where beads of diameter $\sigma=2.5$~nm are connected by finitely extensible non-linear elastic (FENE) springs, and interact via a purely repulsive potential [the Weeks-Chandler-Andersen (WCA) potential --- see below] which captures steric effects. A Kratky-Porod potential introduces a bending stiffness such that the polymer has a persistence length $l_p=20\sigma\approx50$~nm. As in Ref.~\cite{SBrackley2012} we model searcher proteins as a pair of rigidly connected spheres, of diameter $3\sigma$ and $\sigma$ respectively [Fig.~1 in main text]; the larger sphere has only a steric interaction with the DNA, whereas the smaller sphere interacts via a truncated LJ potential such that it is sticky and acts as a DNA binding site. The shifted LJ potential between spheres $i$ and $j$ is given by
\begin{align}\label{eq:LJ}
U_{\rm LJ~shift}(r_{ij})= \left\{ 
\begin{array}{ll} 
U_{LJ}(r_{ij})-U_{LJ}(r_{\rm cut}) & r\leq r_{\rm cut} \\
0 & r>r_{\rm cut}
\end{array} \right.
\end{align}
where $r_{ij}$ is the separation of the beads, $r_{\rm cut}$ is a cut-off distance, and 
\[
U_{LJ}(r)=4 \epsilon' \left[ \left( \frac{d_{ij}}{r} \right)^{12} - \left( \frac{d_{ij}}{r} \right)^{6}  \right],
\]
where $d_{ij}$ is the mean of the diameters of the spheres. Here $\epsilon'$ is an energy scale, but due to the shifting this is not the energy value at the minimum of the potential; for clarity throughout the paper we state interaction energies at the minimum, and call this $\epsilon$. A choice of $r_{\rm cut}=2^{1/6}d_{ij}$ in Eq.~(\ref{eq:LJ}) gives the WCA potential for steric interaction only, and a larger value of $r_{\rm cut}$ gives an attractive interaction (we use a cut off which corresponds to an interaction volume which extends 1~nm from the surface of the sphere --- similar to the Debye length at physiological salt concentrations). 
 Crowders are modelled as single spheres of diameter $2\sigma$ (a typical size for \emph{E. coli} proteins~\cite{SMarenduzzo2006}), and interact only sterically with the DNA. Blockers are modelled as two rigidly connected spheres, this time of diameters $3\sigma$ and $4\sigma$. The larger of the spheres is sticky for the DNA, with a strong interaction of $15~k_BT$. This choice means that the blockers diffuse slowly along the DNA contour, and only occasionally escape from it --- the blocker diffusion constant is $D_{\rm 1D}=0.03~\sigma^2~\mbox{tu}^{-1}$ and they unbind from the DNA with a rate $k_{\rm unbind}=5\times10^{-4}~\mbox{tu}^{-1}$, compared to e.g. a searcher with $\epsilon=4.31~k_BT$ which has $D_{\rm 1D}=0.08~\sigma^2~\mbox{tu}^{-1}$ and $k_{\rm unbind}=1.7\times10^{-2}~\mbox{tu}^{-1}$ (here $\mbox{tu}$ are simulation time units which we define below). By allowing the DNA beads to partially move inside the larger sphere the protein effectively then sits \emph{around} the DNA, blocking the path of the searchers from all sides. By using a two-sphere model we also prevent the blockers from interacting with multiple DNA strands at the same time, for example forming DNA bridges.

Denoting the position of the $i$th sphere by $\mathbf{x}_i$, its dynamics obey the following Langevin equation
\begin{equation}
m_i \frac{d^2\mathbf{x}_i}{dt^2}= -\gamma_i \frac{d\mathbf{x}_i}{dt} - \nabla_i U + \sqrt{2k_BT \gamma_i}\boldsymbol{\xi}_i(t) \label{eq:langevin}
\end{equation}
where $U$ refers to the full potential including the FENE, Kratky-Porod and LJ terms, and $m_i$ and $\gamma_i$ denote respectively the mass and friction experienced by the $i$th particle. $T$ is the temperature, $k_B$ the Boltzmann constant, $\nabla_i=\partial / \partial \mathbf{x}_i$, and $\boldsymbol{\xi}_i(t)$ is an uncorrelated Gaussian noise with zero mean and variance $\langle \boldsymbol{\xi}_i(t)\cdot \boldsymbol{\xi}_j(t')\rangle=\delta_{ij}\delta(t-t')$. We fix the velocity relaxation time $m_i/\gamma_i=1$, and then choose $\gamma_i$ and $m_i$ such that an isolated particle diffuses like a sphere in a fluid of viscosity $\eta=1/3\pi$ according to the usual Stokes-Einstein relation $\gamma_i=3\pi\eta d_i$, where $d_i$ is the diameter of the particle (i.e. for a DNA bead $m_i=1,\gamma_i=1$). 
To map time scales to physical units we define the simulation time unit ${\mbox{tu}=\sigma^2(3\pi\eta\sigma)/k_BT}$, and note that if $\eta=1~\mathrm{cP}$, then one simulation time unit corresponds to $36~\mathrm{ns}$. Systems are equilibrated for at least $10^5$ time steps before the attractive interactions are switched on, so the initial DNA configuration is equilibrated and protein positions are random. We also note that simulation times were in general an order of magnitude larger than e.g. the time needed for a crowder to diffuse across the system, so that any residual influence of the initial condition is lost.  
Hydrodynamic interactions, which would vastly increase the computational cost, are omitted. 

All simulations were performed using the LAMMPS code~\cite{Slammps}. Throughout the paper we use simulation time units, and plots show results averaged over 500 independent simulations, each considering a DNA molecule of length $L=500~\sigma$ in a volume $V\approx50000~\sigma^3$ (leading to a 1\% volume fraction of DNA --- similar to that in an \emph{E. coli} cell). We use a cubic simulation box with periodic boundary conditions. A DNA molecule of this length has a gyration radius of the order $100~\sigma$, meaning that the DNA is compacted (although not quite as much as in a real cell --- we have $R_g/V^{1/3}\sim3$, compared to $R_g/V^{1/3}\sim7$ for \emph{E. coli}). Previous work~\cite{SMirny2009} has shown that, althought there is little effect on overall search time, a compact globule DNA conformation can lead to a decrease in the probability of 3D excursions which start and end at positions with large separation along the DNA; the DNA molecules here are not long enough for this effect to become important. The choice of periodic boundaries ensures that even though the width of the system is only about twice the persistence length, no unrealistic sharp bends are induced in the DNA due to confining walls.

Each simulation is run until the searcher protein binds to the target DNA bead (the central bead in the DNA molecule) for the first time, and the search time and number of rounds is recorded. The 1D and 3D search times are averaged over each search round, and we compute $l_s$ as the number of distinct DNA beads visited during one round of 1D diffusion along DNA (these may be adjacent beads along the DNA contour, or beads which are visited via intersegmental transfer).

\section*{Multiple searcher proteins}

\begin{figure}[t]
\includegraphics{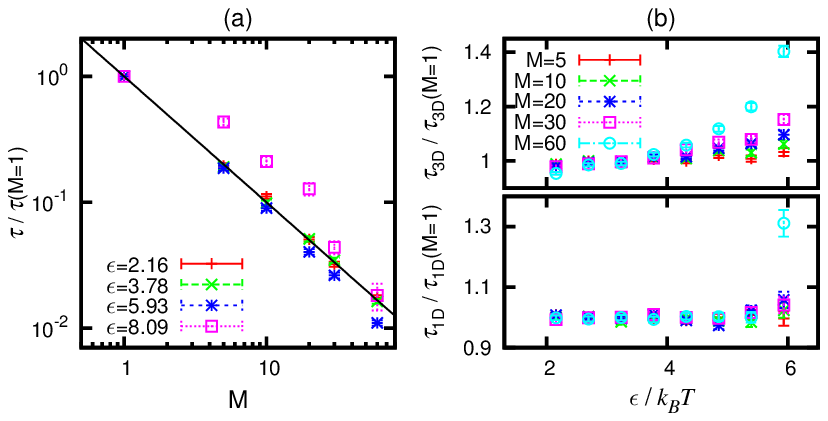}
\caption{Simulation results for systems with multiple searcher proteins. (a) The mean time until the first of $M$ searchers finds a target, as a function of $M$. Results for several different values of $\epsilon$ (units of $k_BT$) are shown, and times are scaled by the search time for a single searcher. The solid line shows $M^{-1}$. (b) The mean 3D and 1D search times as a function of $\epsilon$, for different values of $M$, again scaled by the time for a single searcher.  \label{fig:searchers} }
\end{figure}

In a bacterial cell there may be $\mathcal{O}(10)$ copies of a given transcription factor (e.g. there are typically 5 \emph{lac} repressor tetramers in an \emph{E. coli} cell~\cite{Salberts}). If the search for the target were a simple Poisson process, one would expect that the mean search time should scale inversely with the number of searchers, i.e. the search time for $M$ searchers should be $\tau(M)=\tau(1)M^{-1}$.  Figure~\ref{fig:searchers}(a) shows the search time as a function of $M$ for a number of different protein-DNA affinities, and we see that the $\tau\sim M^{-1}$ scaling holds for $M$ up to around 10 and $\epsilon<8~k_BT$. This result may appear to contradict recent theoretical work by Sokolov et al. \cite{SSokolov2005,SSokolov2005-2}, which predicted an $M^{-2}$ scaling for pure 1D diffusion (which we might expect to be similar to the large $\epsilon$ regime). However, that study applied only in the thermodynamic limit; lattice based simulations of pure 1D diffusion for finite $L$ show an exponent between $-1$ and $-2$ for a finite lattice with $M<10$ (not shown).

In Fig.~\ref{fig:searchers}(b) we show separately the mean duration of the 1D and 3D search episodes, $\tau_{{\rm 1D}}$ and $\tau_{{\rm 3D}}$ respectively. There is a significant dependence of $\tau_{{\rm 3D}}$ on the number of searcher proteins due to steric effects; for small $\epsilon$, $\tau_{{\rm 3D}}(M)\approx\tau_{\rm 3D}(1)$, but for large $\epsilon$, $\tau_{{\rm 3D}}(M)>\tau_{\rm 3D}(1)$. The latter result is due to the large number of searchers bound to the DNA in this case: much of the DNA is occluded by proteins, so it takes longer for a given searcher to encounter a segment of DNA which is free.

Increasing the number of searchers has little effect on the 1D search time, except at large $\epsilon$. When both $M$ and $\epsilon$ are large it is possible for the searchers to form multiple contacts with different DNA segments and become trapped. Some transcription factors (although a minority) have more than one binding site (e.g. the \emph{lac} repressor), and it is only to these that our data for large $\epsilon$ apply. For an isolated searcher protein such multiple contacts are very short lived because the lowering of free energy due to the simultaneous interaction with two DNA strands does not outweigh the entropic cost of bringing the two strands close together; such multiple contacts could be thought to model intersegmental transfers. For large $\epsilon$, if multiple searchers bridge the same DNA segments, the structure is stabilised and the searchers become trapped, leading to the observed increase in $\tau_{\rm 1D}$. Although here we have considered more copies of the same searcher protein than would typically occur in a single cell, the results for $\tau_{\rm 3D}$ and $\tau_{\rm 1D}$ would also apply in the case of many different species of searcher protein. 

Finally we note that there may be some additional advantages to having multiple searchers which we have not considered. We have modelled a uniform DNA molecule, whereas in reality some ``non-target'' DNA sequences might have a very high affinity for protein binding and could act as traps \cite{SBrackley2012} for proteins (although whether the proteins experience a varying potential as they slide remains controversial \cite{SSlutsky2004,SGerland2002,SBenichou2009,SSheinman2012}). If there are multiple copies of a protein, then any falling into such a trap, will block others from entering it.

\section*{Crowders and Blockers - Further Results}

\begin{figure}[t]
\includegraphics{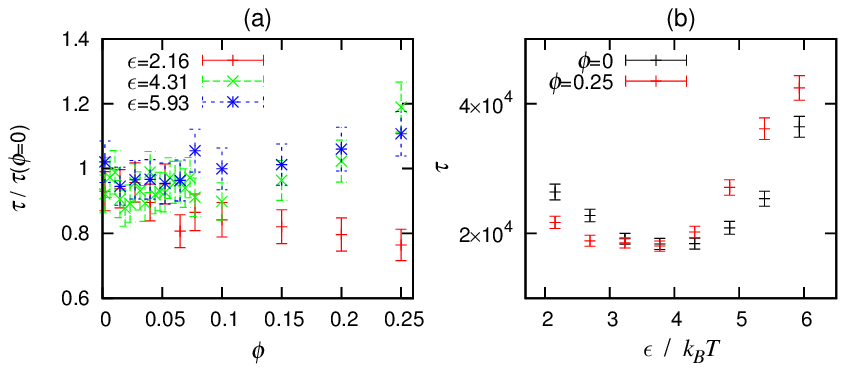}
\caption{Further simulation results for systems with multiple crowder proteins. (a) Search time scale by search time for no crowders, as a function of the crowder volume fraction, $\phi$. (b)~Search time as a function of protein-DNA interaction energy $\epsilon$, for no crowders, and 25\% volume fraction of crowders. \label{crowtau} }
\end{figure}

In Fig.~\ref{crowtau}(a) we show the data from Fig.~3(d), but here the search time is rescaled by the search time observed in the absence of crowders. As detailed in the main text, despite a large change in the number of crowders, there is a relatively small change in search time --- at a 25\% volume fraction the change in search time is around 20\%, despite for example a $\sim50\%$ decrease in $\tau_{\rm 3D}$ and a $\sim70\%$ increase in $\tau_{\rm 1D}$. This is due to the partial cancelling of effects, where $\tau_{\rm 3D}$ decreases, and $\tau_{\rm 1D}$ increases. Also the fact that $\tau$ either increases or decreases depending on the value of $\epsilon$ can be easily seen in this plot.

Figure~\ref{crowtau}(b) shows how the search time varies with $\epsilon$ for the cases of no crowders, and 25\% volume fraction. As detailed in the text, the overall effect of the crowders is to reduce the dependence of $\epsilon$ at the minimum --- there is very little vaiation in $\tau$ for $\epsilon$ between $2.5$--$4~k_BT$. 

\begin{figure}[t]
\includegraphics{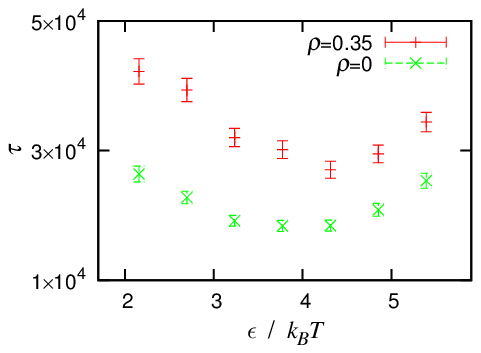}
\caption{Further simulation results for systems with multiple blocker proteins. Search time shown as a function of protein-DNA interaction energy $\epsilon$, for no blockers, and for a blocker DNA coverage density of $\rho=0.35$. \label{blocktau} }
\end{figure}

Figure~\ref{blocktau} shows how the search time varies with $\epsilon$ for the case of no blockers, and of a blocker DNA coverage fraction of $\rho=0.35$. The presence of the blockers results in an overall increase in search time, and this increase is slightly larger for smaller $\epsilon$. The search is dominated by 3D excursions at small $\epsilon$, so the latter observation is consistent with the fact that blockers lead to an increase in $\tau_{\rm 3D}$ [Fig.~4(b) in the main text].

\section*{Correlation between successive slides}

As noted in the text, one of the key assumptions of many models of facilitated diffusion (for example in Refs.~\cite{SHalford2004,SBerg1981,SvonHippel1989}), is that the section of the DNA explored by a protein during each 1D diffusion episode is uncorrelated with that during previous episodes. This leads to the relationship $N_s=L/l_s$ for the average number of rounds of 1D plus 3D diffusion required to find the target. We find in our simulations that this relationship is reasonable when only a searcher protein is present, but breaks down when either crowders or blockers are present. This is demonstrated by a plot of $N_s l_s/L$, which according to the equation above should have a constant value of unity. As detailed in the main text, $l_s$ is measured from the simulations as the mean number of distinct DNA beads visited during a search round. A recent theory developed by Cherstvy et al.~\cite{Scherstvy,Skolomeisky} accounts for correlations in successive 1D searches, but does not take into account steric interactions with crowding proteins.

\begin{figure}[t]
\includegraphics{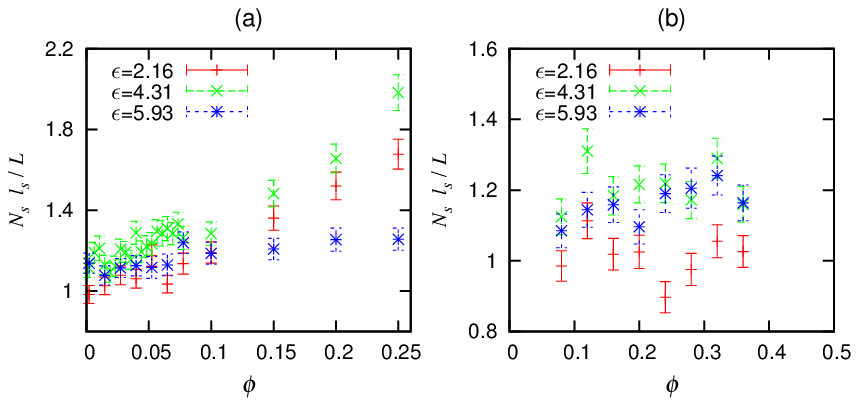}
\caption{Results showing how macromolecular crowding affects the number of search rounds. The quantity $N_sl_s/L$ for different numbers of (a) crowders and (b) DNA bound blockers. Deviation from $N_sl_s/L=1$ indicates a breakdown of key assumptions in classic theories of facilitated diffusion [Eq. (1) of main text]. \label{Nsfig} }
\end{figure}

Figure~\ref{Nsfig}(a) shows that there is an increase in $N_s l_s/L$ with crowder density. This shows that, increasingly with $\phi$, more than $L/l_s$ search rounds are required to find the target. This is consistent with the notion that crowders increase the likelihood that the same region of the DNA is searched during consecutive 1D episodes (since depletion-like effects favor rebinding of a protein soon after it escapes from the DNA). 

In Fig~\ref{Nsfig}(b), it is shown that for the case of blockers there is a more modest increase in $N_s l_s/L$ (except for small values of $\epsilon$, where $N_s$ is anyway much larger than in the other cases). The cause here is likely as noted in Ref.~\cite{SLi2009}, that as the proportion of the DNA which is covered by blockers increases, so too does the probability that the target is also covered. 

\begin{thebibliography}{36}
\bibitem{fn:diffusion}
The motion of the searcher when it is not bound to the DNA remains diffusive for all values of $\phi$ investigated, although its diffusion coefficient decreases with $\phi$~\cite{verkman}, as does that of the DNA beads.
\bibitem{fninterseg}
An intersegmental transfer is an event where a searcher moves from one DNA strand directly to another --- which is close in space, but may be distant along the contour --- by transiently binding with both.
\end{thebibliography}
\end{document}